\documentclass[12pt]{iopart}
\usepackage{iopams}  
\usepackage{graphicx}
\begin{document}

\title[]{Role of triaxiality in the ground state shape of
neutron rich Yb, Hf, W, Os, and Pt isotopes}

\author{L M Robledo$^{1}$, R Rodr\'{\i}guez-Guzm\'an$^{2}$ 
and P Sarriguren$^{3}$}

\address{
$^{1}$ Departamento  de F\'{\i}sica Te\'orica C-XI,
Universidad Aut\'onoma de Madrid, 28049-Madrid, Spain}
\address{
$^{2}$ Department of Physics, P.O.Box 35 (YFL), FI-40014 University of Jyv\"askyl\"a, Finland}

\address{
$^{3}$ Instituto de Estructura de la Materia, CSIC, Serrano
123, E-28006 Madrid, Spain
}

\ead{luis.robledo@uam.es}

\begin{abstract} 
The evolution of the ground-state shape along the triaxial landscape of
several isotopes of Yb, Hf, W, Os, and Pt is analyzed using the
self-consistent Hartree-Fock-Bogoliubov  approximation. Two well reputed
interactions (Gogny D1S and Skyrme SLy4) have been used in the study in
order to asses to which extent the results are independent of the details
of the effective interaction. A large number of even-even nuclei, with neutron
numbers from $N=110$ up to $N=122$ has been considered, covering in this
way a vast extension of the nuclear landscape where signatures of
oblate-prolate shape transitions have already manifested both theoretically
and experimentally.
\end{abstract}

\pacs{21.60.Jz, 27.70.+q, 27.80.+w}

\maketitle

\section{Introduction}
\label{Intro}

One of the most often encountered characteristics of the atomic 
nucleus is the existence
of an intrinsic deformed ground state. Deformation is a direct consequence of the
spontaneous rotational symmetry breaking mechanism of the mean field approximation
and owes its popularity to its ability to incorporate correlations into
the mean field wave function \cite{Bender-Review}. Both, experimental results and theoretical calculations
lead to the conclusion that most of the deformed nuclei show a quadrupole
deformation of the prolate kind (cigar-like shape) that preserves to a
great extent axial symmetry (i.e., there exists a symmetry axis in the matter
distribution). Therefore, those regions of the nuclide chart showing
oblate deformation or deformed mass distributions breaking axial symmetry
(referred to as triaxial distributions) are of great interest to deepen into
the understanding of the shell structure underlying the appearance of
deformation. In this respect, a region of interest is the one with mass
number A around 190 where a prolate to oblate shape transition as a function
of neutron number has been predicted \cite{Bengtsson.87,naza90,Wood-Review} 
as well as some examples of triaxial
ground states. This has fostered both theoretical 
\cite{Naik.04,Stevenson.05,Fossion.06,walker2006,Jolie.03,Sarriguren.08,Sun.08,Morales.08,Garcia-Ramos.09} 
and experimental \cite{wu96,Stuchbery.96,Wheldon.00,podolyak,Caamano.05,Podolyak.09,Pfutzner.02} 
studies in the region. Conclusive experimental results are scarce as it is not 
easy to find an observable sensitive to the sign of deformation and/or
triaxiality that is, at the same time, easy to measure. As a consequence,,Morales.08
theoretical predictions are important in spite of their uncertainties with
related to in-medium effective interactions and/or theoretical methods used to solve 
the problem. 

In the A=190 mass region there has been a variety of theoretical 
calculations in the past mainly using the mean field approach and a variety
of interactions. Our interest in this paper is to investigate the role of
the triaxial degree of freedom in this region emphasizing those features which
are independent of the mean field effective interaction used. To this end, we 
use the Hartree-Fock-Bogoliubov
(HFB) method \cite{RS.80} together with some of the best effective 
interactions/functionals present in the market, namely Gogny D1S \cite{Decharge.80,d1s}
Gogny D1N \cite{Chappert.08} and Skyrme SLy4 \cite{Sly4} to carry out constrained 
calculations in the collective $\beta$ and $\gamma$ deformation
variables in order to obtain the so-called $\beta - \gamma$ planes (potential
energy surfaces as a function of the $\beta$ and $\gamma$ parameters) for the
chemical species $_{70}$Yb, $_{72}$Hf, $_{74}$W, $_{76}$Os and $_{78}$Pt
and neutron numbers from N=110 until 122 in steps of two units. In this work
we study both the ground state shape evolution as the number of neutrons increases and
the role of triaxiality in these isotopes. The transition from axially symmetric
prolate shapes to axially symmetric oblate shapes passing through $\gamma$-soft
triaxial nuclei could illustrate good examples of the transition from the SU(3)
dynamic symmetry of the interacting boson model (IBM) \cite{Iachello.87} to the
$\overline{\textrm{SU}(3)}$ symmetry passing across the O(6) dynamic symmetry describing
$\gamma$-soft systems. In Section 2 the
relevant technical details of the calculation, definitions of the quantities used
and a description of the interactions/functionals used is given. In Section
3 we present results for the $^{190}$W and the three interactions/functionals
used. Once the equivalence of the results is stated, the deformation systematics
is analyzed by using results with D1S and SLy4. In order to get some insight
on the relevant configurations we have also discussed in this region the
single particle energies (SPE) both along the axial and the triaxial degrees of
freedom in the paradigmatic $^{190}$W case. By using this single particle
plots we get an overall understanding of the evolution of deformation in
this region. We end up this section by comparing the selfconsistent moments
of inertia obtained with the experimental results. Finally, in Section 4 the
conclusions are presented.

\section{Solution of the mean field equation and interactions used}

\subsection{Mean field equation and its solution}

To obtain the mean field wave functions we treat pairing correlations in the 
framework of the HFB approximation \cite{RS.80}. 
Taking into account that our aim is to study triaxiality, breaking 
of axial symmetry is allowed in the numerical procedure to solve the 
HFB equation. On the other hand, the discrete symmetries parity, time 
reversal, and simplex are preserved in the calculation. Keeping parity as a 
good quantum number is not a severe 
constraint as octupole deformation effects are not expected to be relevant
in the region under study. Preserving time-reversal restricts the treatment
to even-even nuclei and zero spin (i.e., the ground state). Finally, simplex
is a standard symmetry preserved in almost all the HFB calculations performed
up to now \cite{Goodman.77} as it is supposed to play essentially no role in the dynamics of
the ground state of atomic nuclei. Besides the usual constraint on the
average number of protons and neutrons, which is characteristic of the HFB approximation,
we have constrained the mean value of the quadrupole operators
$Q_{20}=z^2-\frac{1}{2}(x^2+y^2)$ and $Q_{22}=x^2-y^2$, as a way to obtain
the standard $\beta-\gamma$ plane of any triaxial study. Instead of the
$\beta-\gamma$ plane we will plot the $Q_0-\gamma$ plane where the deformation
parameter $\beta = \sqrt{4\pi/5} Q_{20} /(A \langle r^2 \rangle )$ is 
replaced by 
$$
Q_0=\sqrt{Q_{20}^2+Q_{22}^2}.
$$
The $\gamma$ angle is defined as usual as $\tan \gamma =Q_{22}/Q_{20}$. 
With this definition an axially symmetric prolate mass distribution has a $\gamma=0 ^\circ$ value
whereas the corresponding oblate has $\gamma=60 ^\circ$. 

The single particle energies $\epsilon_k$ whose evolution as a 
function of both $Q_{20}$ and $\gamma$ degrees of freedom is shown and 
discussed in length in the next section are obtained as the eigenvalues 
of the Hartree-Fock Routhian $h'=t+\Gamma - \lambda_2 Q_{20} -\lambda_{22} Q_{22}$, 
where $t$ is the kinetic energy operator, $\Gamma$ is the Hartree-Fock field 
and $\lambda_2 Q_{20}+\lambda_{22} Q_{22}$  represents the standard 
Lagrange multiplier term used to enforce the constraint on the mean values
of the $Q_{20}$ and $Q_{22}$ operators. As the
HF Routhian preserves parity the single particle energies are labeled with
the parity quantum number. Obviously, for axially symmetric
shapes the last Lagrange multiplier term is missing and also the quantum numbers
labeling the SPE include, in addition to parity, the third component $K$ of
the intrinsic angular momentum operator along the $z$ direction.  
Also, due to the Coulomb energy and the different
number of protons and neutrons, the SPE for each 
kind of nucleon are different and will be shown separately. 
Due to time reversal (in the axial case) and simplex (in the
triaxial case) invariance imposed in the calculations, the 
single particle energies are doubly degenerate.
It is also worth pointing out that the SPE have no
direct physical meaning in the framework of the HFB method but they closely
resemble what would be obtained by performing a pure HF calculation and
therefore are useful quantities when the physics is explained
in terms of arguments concerning level densities. To create continuous lines
the non-crossing rule that inhibits the crossing of levels with the
same quantum numbers ($K$ and parity in the case of the plots corresponding
to axially symmetric configurations, and parity alone in the triaxial case) has been used.

As it will be discussed in depth later, we have performed calculations with two kinds of
interactions, namely the Gogny force \cite{Decharge.80} (D1S \cite{d1s} and D1N \cite{Chappert.08}) 
and the Skyrme functional (SLy4) \cite{Sly4} in the particle-hole channel 
plus a zero range and density dependent interaction \cite{terasaki} in the particle-particle 
channel.
Depending on the interaction different approaches to solve the HFB equation have been
used. In the case of the Gogny force, the quasiparticle operators have been 
expanded in a Harmonic Oscillator (HO) basis big enough (thirteen shells) 
as to guarantee the convergence of the observable quantities. 

The solution of the HFB equation in the case of the Gogny force has been obtained 
by expressing the problem as a minimization process on the mean field energy. With 
this in mind, the Thouless parametrization \cite{RS.80} of the most
general HFB wave function has been used to express the HFB energy
as a function of the Thouless parameters. The ones corresponding to the
solution of the HFB problem are obtained by minimizing the energy using
standard gradient methods \cite{Egido.95}. The advantage of this method
of solution is that the implementation of many constraints (as it is needed 
in the present calculations) is straightforward and very easy to implement
in a computer code as it only involves imposing orthogonality of certain 
vectors. As it is customary in calculations
with the Gogny force \cite{Decharge.80}, the two body center of mass kinetic energy correction has
been fully taken into account in the minimization process. Concerning the
Coulomb interaction, its contribution to the direct mean field potential
is fully taken into account. On the other hand, the Coulomb exchange energy
is treated in the Slater approximation and the contribution of the Coulomb
interaction to the pairing field is completely neglected. 

In the case of the Skyrme HF+BCS calculations our main tool has been the code 
EV8 \cite{ev8} and we have taken full advantage of its three-dimensional 
Cartesian lattice discretization \cite{Bender-Review,ev8} to search for general triaxial
solutions. The method used in this 
code to solve the HF+BCS equations is  the successive iterations one 
that relies on an iterative diagonalization of the HF+BCS hamiltonian. 
For details the reader is referred to Refs. \cite{Bender-Review,ev8} and 
for a recent application of this scheme to study both axial and triaxial 
ground state shapes is referred to Ref. \cite{Sarriguren.08}.

\subsection{Interactions and functionals}

In the case of the Gogny force \cite{Decharge.80}, two different parametrizations have been used, namely 
the D1S \cite{d1s} and D1N \cite{Chappert.08} parameter  sets. The former was adjusted more than 30 years 
ago in order to reproduce several nuclear matter properties of interest as well as  some characteristics of 
selected spherical nuclei. Finally, a reasonable surface energy was chosen in order to reproduce the 
fission barrier heights of the actinides. On the other hand, the D1N parameter set has been recently 
proposed with the twofold aim of having a better reproduction of the equation of state of neutron matter 
(as a way to obtain reasonable characteristics in neutron rich nuclei) and reducing the linear trend 
observed in the plots of binding energy differences (theory minus experiment) as a function of neutron 
or proton numbers. In both cases, the central part of the interaction is finite range, what allows to
use it also to obtain the particle-particle pairing interaction in a consistent fashion. The predictive
power of D1S and its ability to reproduce low energy experimental data all over the nuclide
chart are well established (see Refs. \cite{Sarriguren.08,Egido.93,Afanasjev.00,Rodriguez.00,Rodriguez.02,Egido.04,Rodriguez.04}
for some relevant references related to the present discussion). For D1N still many calculations have to be
performed to asses its abilities but it is quite likely that it will also
prove to be a reliable interaction all over the nuclide chart.

Concerning the Skyrme functional (SLy4) it was also fitted \cite{Sly4} to 
reproduce neutron matter properties appropriately and it has proved to give 
reasonable results for many observables all over the nuclide chart. For the
pairing channel we have used a zero-range density-dependent pairing 
interaction (DDPI) \cite{terasaki}, 
\begin{equation}  
V({\bf r_1},{\bf r_2})=-g \left( 1-\hat{P}^{\sigma} \right)
\left( 1-\frac{\rho({\bf r_1} )}{\rho_c}\right)
\delta ( {\bf r_1}- {\bf r_2}  )\, ,
\label{dd-pairing}   
\end{equation}
where $\hat{P}^{\sigma}$ is the spin exchange operator, $\rho({\bf r})$
is the nuclear density, and the parameter $\rho_c=0.16$ fm$^{-3}$. 
The pairing's interaction strength  $g$ is taken as $g=1000$ MeV fm$^3$  
for both neutrons and protons and a smooth cut-off of 5 MeV around the 
Fermi level has been introduced \cite{terasaki,Rigo}.
The motivations for this choice are the very reasonable results obtained with this
combination in systematic studies of correlation energies from $^{16}$O to the 
superheavies \cite{Sly4.res1} and the nice reproduction of experimental
data in global studies of spectroscopic 
properties of the first $2^{+}$ states in even-even nuclei \cite{Sly4.res2}.  
Thus, the predictive power of this combination of effective interactions, 
has been well established along the nuclear chart.

\section{Results}

\subsection{The nucleus $^{190}$W}

The nucleus $^{190}$W corresponds to N=116 and it is therefore in the 
middle of the region of nuclei  studied in this paper. This makes it a 
good candidate for a detailed  explanation of the kind of results obtained for 
other nuclei. 
We have performed calculations for two different parametrizations of the Gogny force 
(the old D1S and the newly postulated D1N) and the Skyrme SLy4 one with 
the DDPI pairing force with strength $g=1000$ MeV fm$^3$ for both protons 
and neutrons.

\begin{figure}
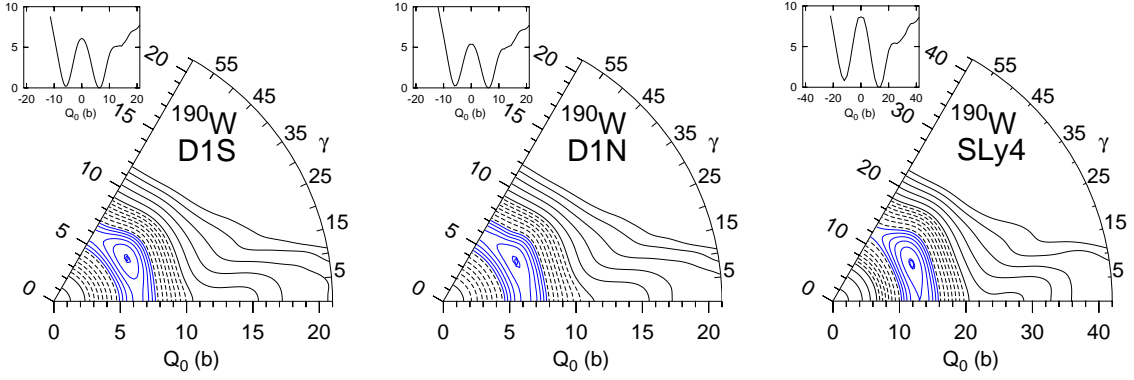

\includegraphics[width=0.33\textwidth]{Fig1a.ps}%
\includegraphics[width=0.33\textwidth]{Fig1b.ps}%
\includegraphics[width=0.33\textwidth]{Fig1c.ps}
\caption{(Color online) $Q_0-\gamma$ planes computed with the two Gogny parameter sets 
used (D1S, left;  D1N middle) as well as with the Skyrme SLy4 
(right) in the nucleus $^{190}$W. The minimum is marked with a small circle. 
The separation between contour lines is of 250 keV for the full line blue contours
around the minima up to 1.5 MeV. It is of 0.5 MeV for the dashed line contours with
energies from 2 up to 4.5 MeV. Finally, the furthest away from the minimum, 
full line contours are separated 1 MeV and span a range of energies
from 5 MeV up to 10 MeV. Note
that the $Q_0$ parameter for SLy4 is defined as twice the $Q_0$ used for the Gogny force 
calculations.}
\label{Fig190W}
\end{figure}

The main results of these calculations are shown in Fig. \ref{Fig190W}. 
There, the potential energy surfaces (PES) in the form of $Q_0-\gamma$  
planes are depicted for the  three interactions/functionals considered. 
In the three cases, the minimum corresponds to a triaxial configuration 
with $\gamma \approx 30 ^\circ $ but with a very small depth with respect 
to the axially symmetric saddle points (i.e., the prolate and oblate 
minima obtained when $\gamma$ is not considered, and that become saddle 
points in the extended parameter space including the $\gamma $ degree of 
freedom as a consequence of the emergence of the triaxial minimum). The depth is
of around 300 keV for the D1S force calculation, it is reduced to around 
100 keV for D1N and goes up again up to around 250 keV in the case of
the Skyrme SLy4 functional.  As a consequence, the axially symmetric
prolate and oblate saddle points/minima are almost degenerate with the
triaxial minimum in the three cases as can
be observed in the small insets depicting the potential energy curves (PEC)
along axially symmetric shapes. Only the SLy4 functional calculation 
shows a somehow higher oblate minimum lying at around 1 MeV above 
the prolate one. It is also worth mentioning that 
the spherical configuration in the SLy4 calculation lies at a higher 
energy as compared to the prolate minimum than in the case of the Gogny force 
calculations. This effect has already been observed in other systematic calculations 
in the same region \cite{Sarriguren.08} and could be due to different pairing
properties of the two forces/functionals. 

We conclude that the Gogny force results using D1S and D1N are very similar
and therefore in the next sections only results with D1S will be presented.
On the other hand, the slight differences observed between the Skyrme
functional and Gogny force results as well as the intrinsic differences
between the two (zero versus finite range, mainly) warrant the comparison of both results
in the subsequent discussions.

\subsection{Deformation systematics}

\begin{figure}
\includegraphics[width=0.85\textwidth]{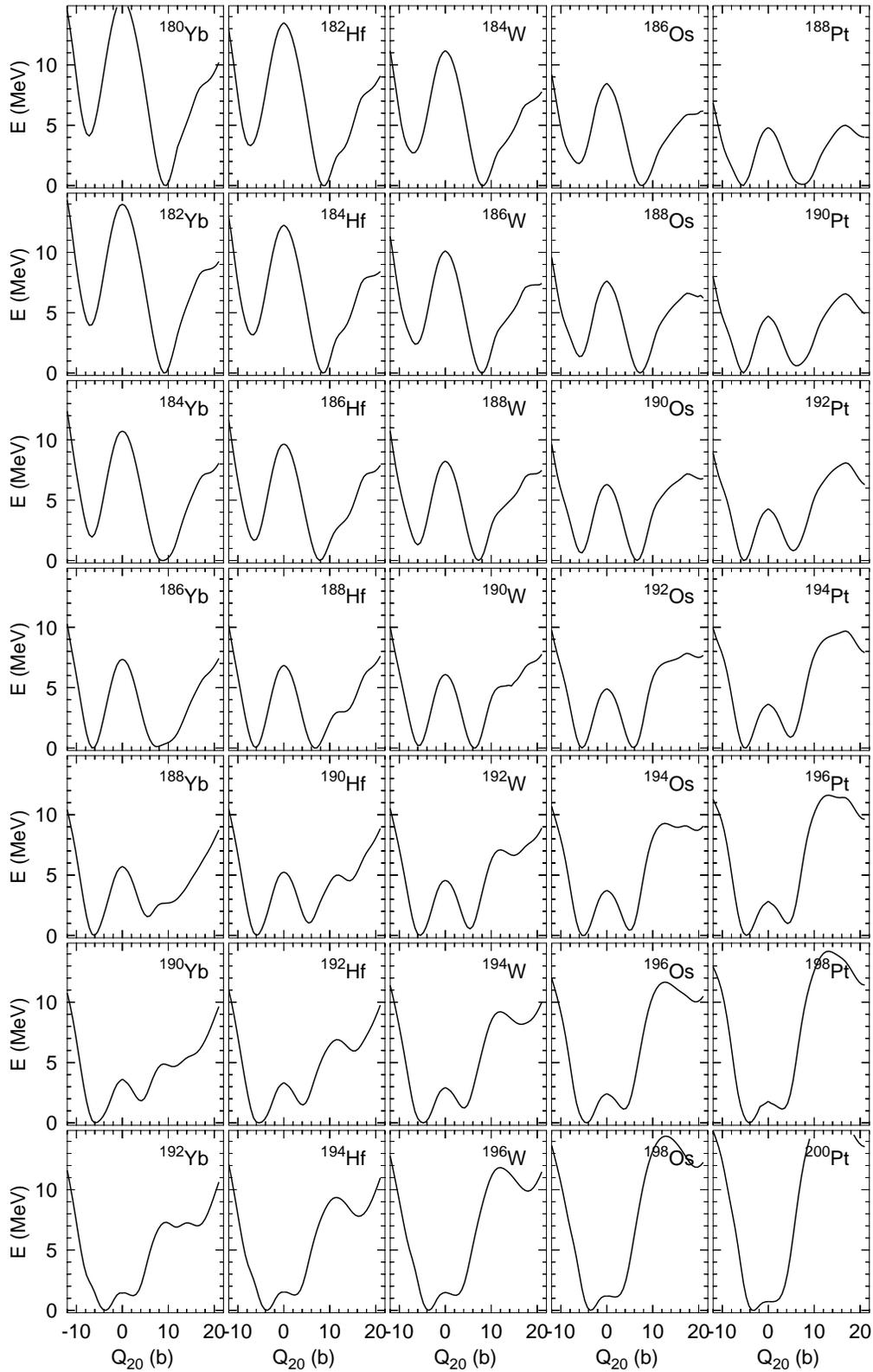}
\caption{Potential energy curves as a function of the axial quadrupole moment $Q_{20}$ computed with
the Gogny D1S interaction for all the nuclei considered. Each row corresponds to a fixed
neutron number ranging from N=110 for the top row up to N=122 for the bottom one.}
\label{PESAX}
\end{figure}

In this section we present the systematics of all the nuclei considered and the 
results obtained with the two interactions/functionals used. First, we show in 
Fig \ref{PESAX} the potential energy curves (PEC's) obtained with the Gogny D1S 
force by constraining on the axially symmetric
quadrupole moment both in the prolate ($Q_{20} > 0$) and oblate ($Q_{20}<0$) side. The
prolate side is equivalent to the triaxial results obtained with $Q_0=Q_{20}$ 
and $\gamma=0^\circ$  whereas the oblate side is equivalent to $Q_0=|Q_{20}|$ 
and $\gamma=60^\circ$. The value of $Q_{20}=10$ b 
roughly corresponds to a $\beta=0.3$ deformation parameter. We observe how in all
the cases there are always a prolate and an oblate minimum even for the N=122 chain
where the prolate minimum is just a mere pocket in the PEC. A naive interpretation
of the presence of the two minima will lead to the conclusion that two rotational
bands, one prolate the other oblate, would be present in the rotational spectra of the
nuclei considered (exception made of some nearly spherical nuclei in the right 
lower corner of the figure). As we will discuss below, the effect of triaxiality 
leads to substantial modifications on the character of many of the observed 
minima converting them into saddle points (see below).
The two minima  lie quite close in energy in many cases (shape coexistence)  an are separated
by a spherical barrier whose height decreases with increasing  Z and N. 
The fact that the coexisting minima lie at more or less the same (in absolute value) 
$Q_{20}$ parameter, suggests the possibility of a triaxial path connecting them as it
is indeed the case (see below). A prolate to oblate transition is observed at N=116. 
This is a very interesting fact, but we defer the discussion 
of this effect until the $\beta-\gamma$ planes have been presented. Also superdeformed structures
can be seen at the highest deformations considered, they are specially relevant (low
excitation energies as compared to the ground state) the higher the Z value and the lower 
the N value of the nucleus are ($^{188}$Pt). These SD structures will not be discussed in the
present paper. Similar results to these ones but for the SLy4 functional have been discussed in 
detail in Ref. \cite{Sarriguren.08}.

\begin{figure}
\includegraphics[width=0.95\textwidth]{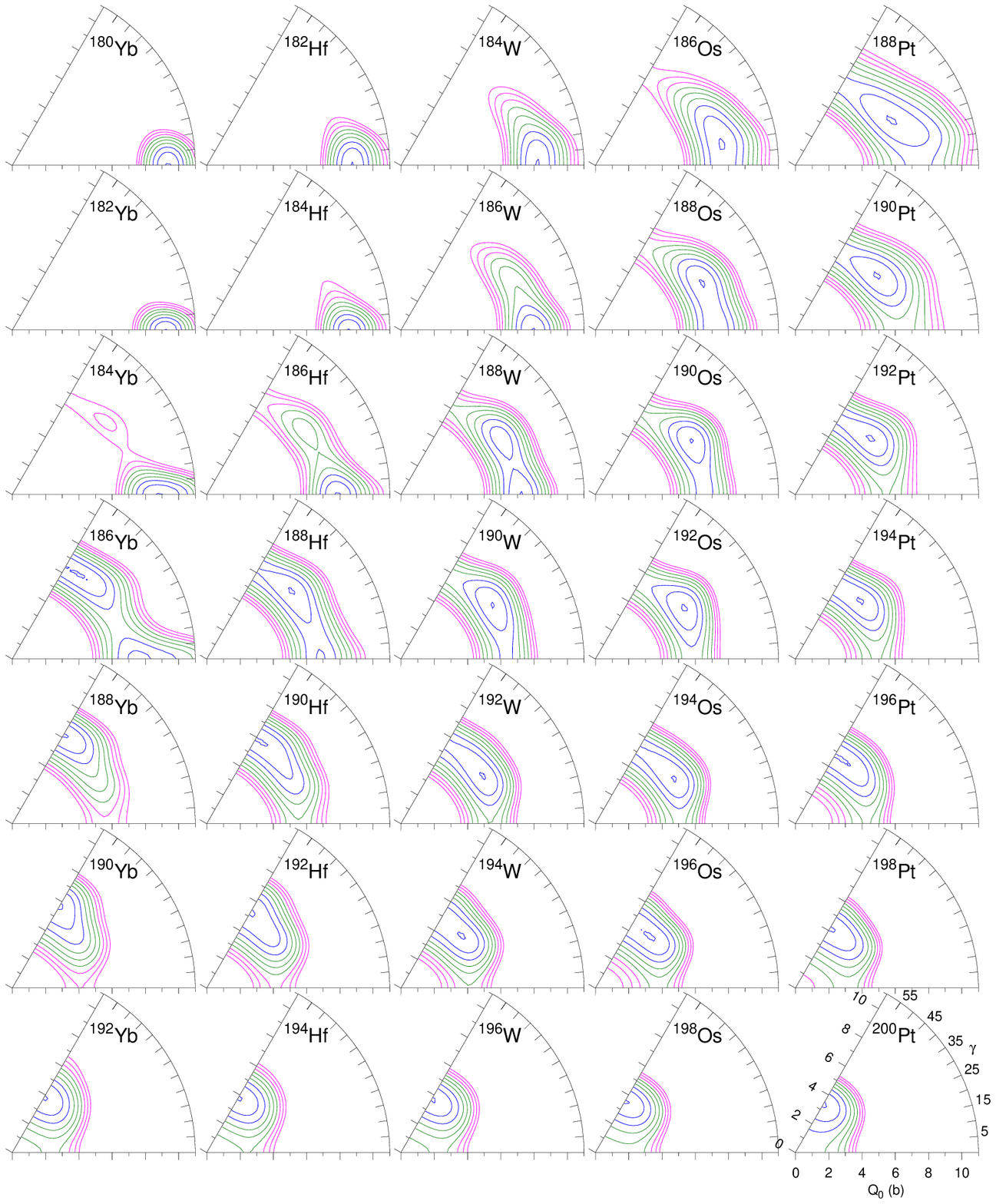}%
\caption{(Color online) $Q_0-\gamma$ planes computed with the Gogny D1S force for all
the isotopes considered. The range of $Q_0$ considered has been reduced
as to focus on the interval around the minima. The contour lines go 
from the minimum energy up to 2 MeV higher in steps of 0.25 MeV. Blue contours
are the three lowest, green ones the next three and magenta contours correspond
to the three with higher energies.}
\label{SysD1S}
\end{figure}

In Fig. \ref{SysD1S} the results of the triaxial calculation and obtained with the Gogny D1S force
are presented. In order to simplify the presentation, in the $Q_0-\gamma$ planes presented the range of 
$Q_0$ is reduced to half the one computed and the number of contour lines considered has also been 
severely reduced by considering contours every 250 keV and up to an 
energy 2 MeV higher than the one of the minimum (which is marked with a small circle).
By looking at this picture several general conclusions can be extracted. The
first one is that increasing Z, for fixed N, drives the corresponding nuclei towards 
triaxiality in such a way that the Pt isotopes (the ones with the highest Z)
are almost all of them triaxial (the exception are $^{198-200}$Pt). Second,
by increasing N for fixed Z, we observe that there is a transition from prolate
to oblate shapes. For N=116 and Z=70 ($^{186}$Yb) there is a sharp transition 
from a prolate ground state
(N $<$ 116) to an oblate one (N $>$ 116). For the neighboring nuclei with Z=72  the
same prolate-oblate shape transition is present but it takes place
in a much broader range of neutron number values involving N=114, 116 and 118 where
the ground state is triaxial. For Z=74 the range of triaxiality extends a
little further away up to N=120. For the higher values of Z (76 and 78 
corresponding to Os and Pt) and N=110 the ground state is already triaxial
and it keeps so up to N=122 for Z=76 and N=120 for Z=78 where it 
becomes oblate. These conclusions are consistent with other theoretical
findings using different interactions 
\cite{naza90,Naik.04,Stevenson.05,Fossion.06,wu96,Wheldon.00}.
Concerning the triaxial minimum we can say that it is in all
the cases very shallow and never reaching a depth of more than 0.5 MeV below 
the saddle points (see below).

\begin{figure}
\includegraphics[width=0.85\textwidth]{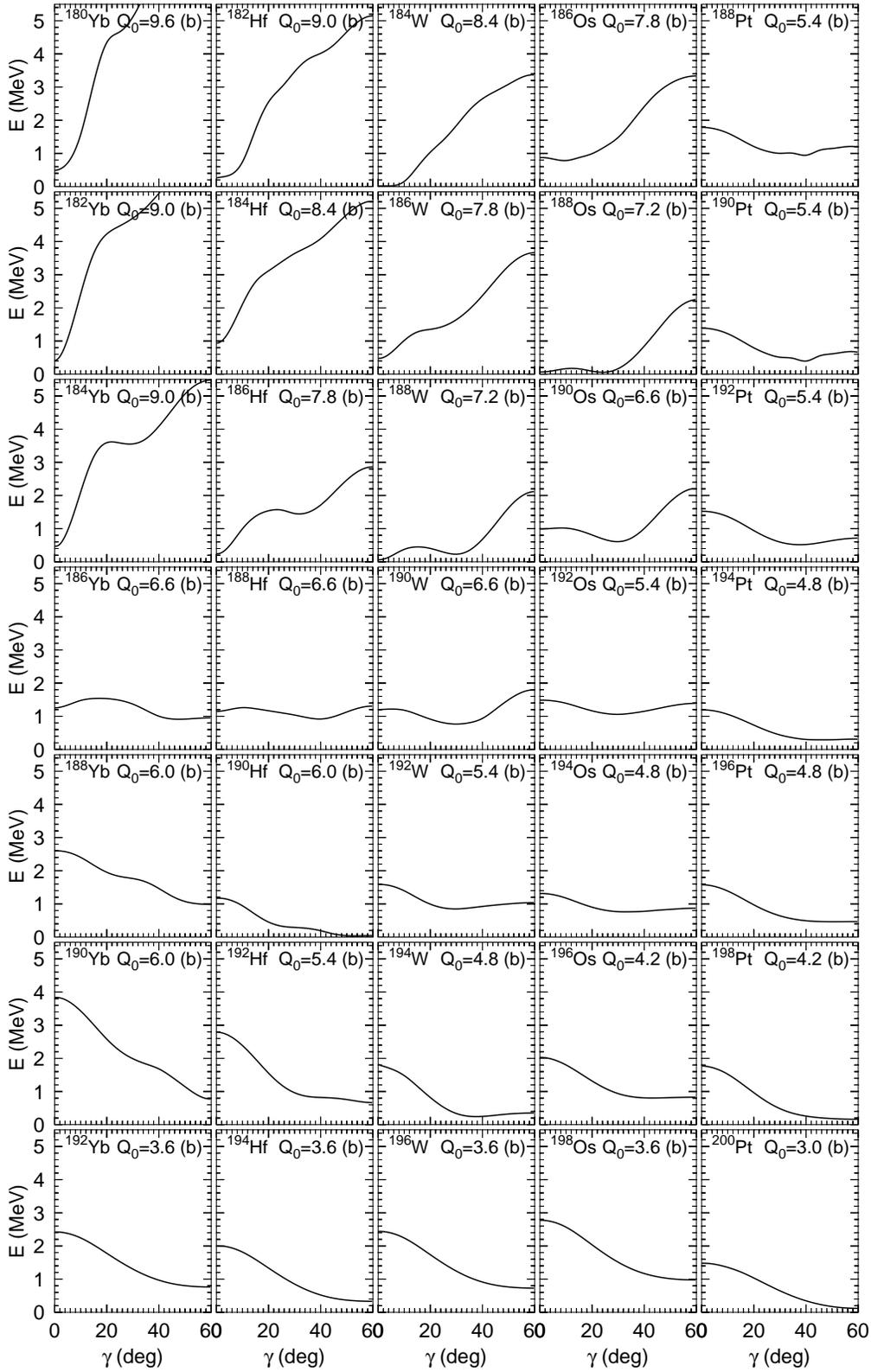}
\caption{Mean field energies computed with the Gogny D1S force are displayed as
a function of the triaxial deformation parameter angle $\gamma$ for fixed values
of $Q_0$ corresponding to the lowest energy of the axially symmetric configurations.}
\label{PESQ0MIN}
\end{figure}

In Fig. \ref{PESQ0MIN} we show the HFB energy as a function of the $\gamma$ deformation parameter for constant
$Q_0$ values (given in each panel) corresponding to the lowest axial minima. This figure is complementary
to Figs. \ref{PESAX} and \ref{SysD1S} and is presented here with the aim of providing a more quantitative understanding of the
PES presented. The most striking conclusion from this figure is that of the two axial minima only one
remains in most of the cases, the other becoming a saddle point. This is manifest in the N=120 and 122
chains where the only remaining minimum is the oblate one ($\gamma=60^\circ$). For N=118 we have a 
similar situation but in this case there are two nuclei with only one very shallow
triaxial minimum ($^{192}$W and and $^{194}$Os). In the N=116 chain we have three shallow triaxial
minima for the Hf, W and Os isotopes and a nucleus, $^{186}$Yb, showing a prolate and oblate minima
but separated by a quite low barrier. For N=114, we have three nuclei (Yb, Hf, and W) with 
prolate and very shallow triaxial minima and the other two with only one triaxial minimum. For N=112 and
110 the Yb, Hf and W nuclei only show a prolate minimum whereas the Os and Pt show very shallow
triaxial minima (and a extremely shallow prolate one in $^{188}$Os). From the above discussion we
can conclude that in most of the cases only one minimum remains, reducing thereby  by half the number of
rotational states to be expected. We can also conclude that due to the shallowness of many minima a
dynamical treatment considering both $Q_0$ and $\gamma$ degrees of freedom will be quite relevant for
a more quantitative understanding of the isotopes discussed.

\begin{figure}
\includegraphics[width=0.95\textwidth]{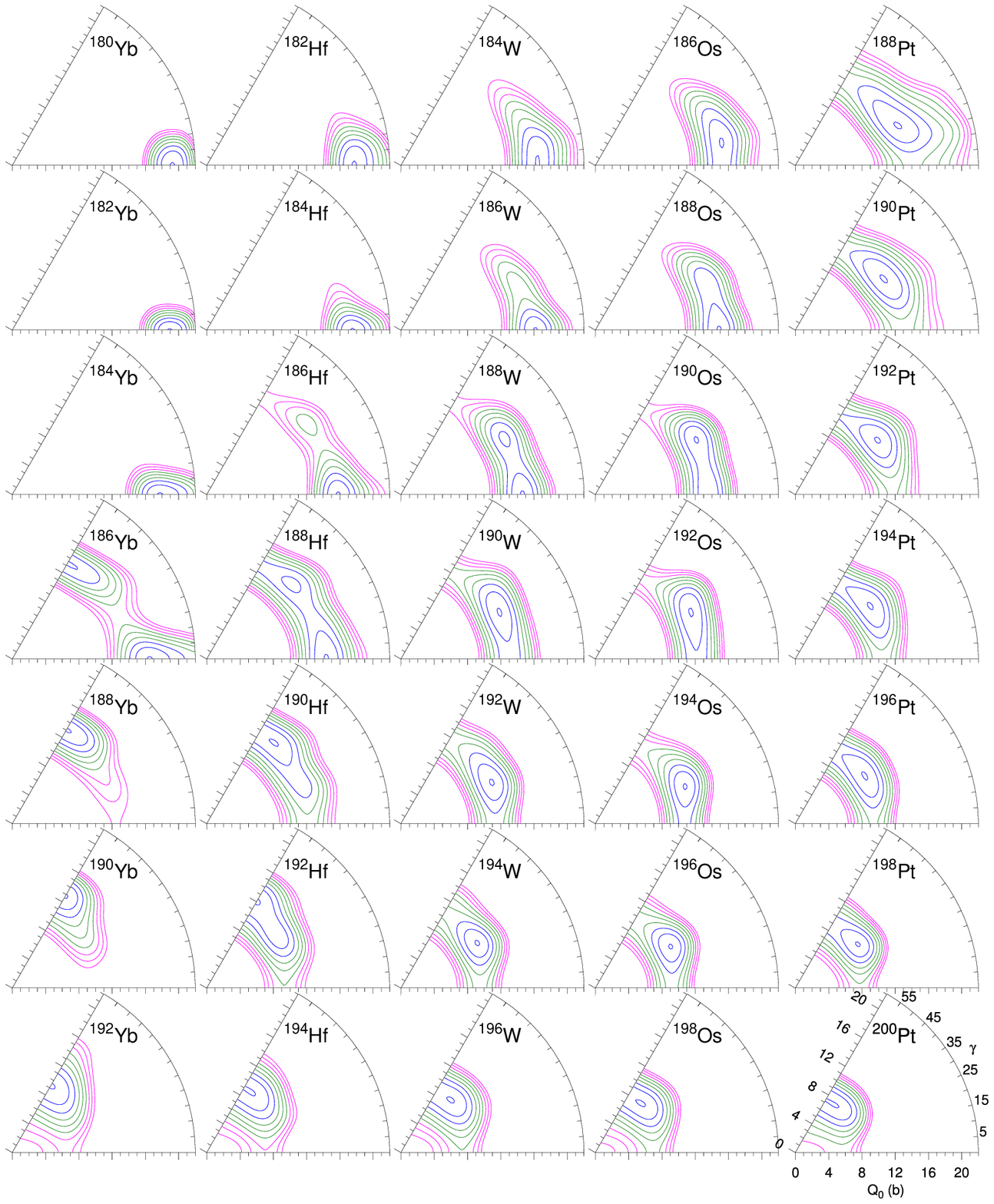}%
\caption{(Color online) Same as in Fig. \ref{SysD1S} but for SLy4 Skyrme functional.}
\label{SysSly4}
\end{figure}

In Fig. \ref{SysSly4} we present the $Q_0-\gamma$ planes computed with
the Skyrme SLy4 functional and for the same nuclei as before. The first
and most relevant fact is that, apart from some details, both pictures (this
and Fig. \ref{SysD1S}) look
rather similar. The prolate-oblate and prolate-triaxial-oblate transitions
show up more or less in the same places in both cases and the contour
plot patterns look rather similar. In principle this fact should not be surprising as 
the big picture of deformation emerges from the interplay between two
bulk properties, namely the surface energy and the Coulomb repulsion. As
both the D1S force and SLy4 functional are adjusted as to carefully 
reproduce bulk nuclear matter properties one could expect a nice agreement
between the deformation related predictions. However, the fine details
of deformation are strongly dependent upon shell effects and pairing 
properties of the interactions and those are definitely not the same in
D1S and in SLy4. As a consequence of those details we notice that the
triaxial minima are typically around 0.8 MeV deeper with SLy4 than with D1S
for nuclei with neutron number greater than 116.
Because of this, we find rather deep triaxial minima (around 1.25 MeV and
more) in nuclei like $^{192}$W and $^{194}$Os. We also notice that the
N=122 nuclei that were all of them oblate for D1S are now triaxial with
SLy4, exception made of $^{192}$Yb, but the depth of the minima never
exceed 0.25 MeV so that a pure triaxial character can not be unambiguously
attributed to those nuclei. 

From the above discussions we can conclude that the prolate to oblate
transition taking place at N=116 as well as the tendency towards triaxial
shapes as proton number Z is increased for fixed N are genuine predictions
as they are present for the two force/functional considered. On the other
hand, and concerning the degree of triaxiality of the properties of the
nuclei showing triaxial minimum the present results are more uncertain as
the depth of the triaxial minima are not deep enough as to make any 
quantitative assertion without consider the dynamics of the relevant degrees
of freedom. It is clear that, for a more quantitative 
description, the fluctuations in the $Q_0$ and $\gamma$ degree of freedom have to
be incorporated as it has been recently been done \cite{Li.09} in other regions of
the periodic table 
in the framework of the five dimensional Bohr hamiltonian. Work along
this direction is in progress and will be reported elsewhere.

To finish this section we have included in Table \ref{TableBetaG}
the numerical values of the $\beta$ and $\gamma$ deformation parameters 
for the ground state solution obtained with the D1S  in a consistent fashion from the same  Gogny force. 
We observe that the ground state 
$\beta$ value decreases as the number of neutrons increase
and at the same time the  $\gamma$ parameter increases. The behavior
of $\beta$ is not surprising because as N increases it comes closer 
to the magic number N=126. The behavior with increasing proton number is
similar, and $\beta$ decreases when Z tends towards the magic value Z=82. 
On the other hand, no specific behavior emerges for the values of the $\gamma$ 
parameter although in general they tend to move from axially deformed to 
$\gamma$ soft. 

\begin{table}
\begin{tabular}{|c|c|c|c|c|c|}\hline
N   & $_{70}$Yb    &   $_{72}$Hf    &    $_{74}$W    &    $_{76}$Os   &    $_{78}$Pt   \\ \hline\hline
110 & (0.28,1.0)  &  (0.26,0.0)  &  (0.24,0.0)  &  (0.23,10.0) &  (0.18,23.8) \\ \hline
112 & (0.28,0.0)  &  (0.25,0.0)  &  (0.23,0.0)  &  (0.20,24.3) &  (0.17,32.6) \\ \hline
114 & (0.26,0.0)  &  (0.23,0.0)  &  (0.21,0.0)  &  (0.19,29.7) &  (0.16,36.2) \\ \hline
116 & (0.19,49.3) &  (0.19,39.5) &  (0.18,29.2) &  (0.17,29.4) &  (0.15,40.4) \\ \hline
118 & (0.18,60.0) &  (0.17,54.0) &  (0.16,28.3) &  (0.15,28.2) &  (0.13,45.0) \\ \hline
120 & (0.17,60.0) &  (0.15,60.0) &  (0.13,37.4) &  (0.12,38.0) &  (0.11,60.0) \\ \hline
122 & (0.10,60.0) &  (0.10,60.0) &  (0.10,60.0) &  (0.09,60.0) &  (0.09,60.0) \\ \hline
\end{tabular}
\caption{Deformation parameters $(\beta,\gamma)$ for the ground state minimum obtained 
with the Gogny D1S interaction.}\label{TableBetaG}
\end{table}

\subsection{Single particle energies}

\begin{figure}
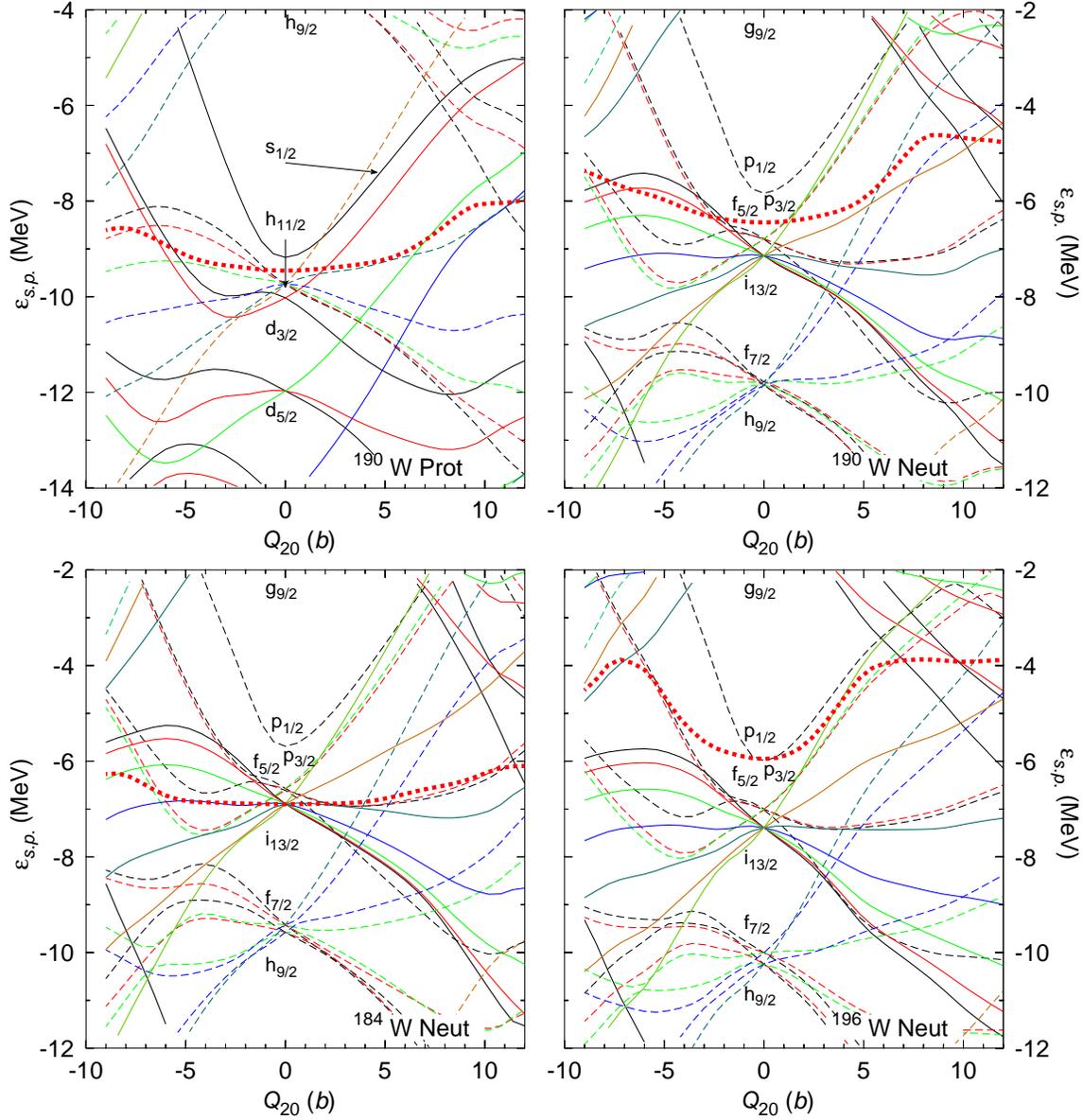

\includegraphics[width=0.5\textwidth,angle=270]{Fig6a.ps}

\includegraphics[width=0.5\textwidth,angle=270]{Fig6b.ps}
\caption{(Color online) Upper panels: Single particle energies for protons 
(left panel) and neutrons (right panel) and the nucleus
$^ {190}$W plotted as a function of the axial quadrupole moment 
$Q_{20}$ for both positive (prolate) and negative (oblate) side. 
The Fermi level is depicted in both cases as a thick dashed red line. 
The results have been obtained with the Gogny D1S force. 
Full line curves correspond to levels with positive parity whereas
dashed lines correspond to negative parity states.  
The color labeling is as follows and with increasing values of 
$K=1/2, 3/2, 5/2, \ldots$, black, red, green, blue, dark-blue, brown, 
dark-green, etc. Lower panel: same as above but for the neutron SPE 
of $^{184}$W (left) and $^{196}$W (right).
}\label{190Wspe}
\end{figure}

Now we turn our attention to the SPE plots 
obtained as a function of the axial quadrupole moment $Q_{20}$  for the 
selected nuclei $^{184}$W, $^ {190}$W and $^{196}$W. 
The SPE plots obtained for other nuclei and/or other interactions (D1N or 
SLy4 functional) are quite similar to the ones depicted here and thus 
we consider only these as representative examples. The election is based
on the fact that  $^{184}$W has a prolate ground state, $^ {190}$W is triaxial 
whereas  $^{196}$W shows a minimum in the oblate side. Then, by looking 
at the SPE we hope to find the features that drive these systems towards
their characteristic deformations.

The SPE are plotted in Fig. \ref{190Wspe} for the three nuclei mentioned.
The SPE are depicted as a function of the axial quadrupole moment $Q_{20}$
and therefore they correspond to axially symmetric configurations (both
prolate and oblate). Below we will consider also the behavior of the SPE
as a function of the triaxial parameter $\gamma$.
Only the proton's SPE of $^{190}$W are plotted as for the other two isotopes
they look very similar to the ones already shown.
The SPE levels for axially symmetric configurations are tagged by the (half integer) 
$K$ quantum number that corresponds to the third component of the 
angular momentum in the intrinsic frame. 
As a consequence of time-reversal invariance, orbitals with
the same absolute value of $K$ are degenerate (Kramers degeneracy) 
and therefore they appear as a single line in the plot.
The levels gather together at $Q_{20}=0$ to form the spherical shell model orbitals with 
quantum numbers $nlj$. The tags of the most relevant shell model orbits
are indicated in the plot. Finally, the (positive) negative parity levels are
plotted as (full) dashed lines. In the plot corresponding to the protons
in $^{190}$W we observe the presence of  the $3s_{1/2}$ level just above 
the Fermi level and below the $1h_{11/2}$, $2d_{3/2}$ and $2d_{5/2}$. 
For neutrons and $^{190}$W we have the $3p_{1/2}$ level 
above the Fermi level and an almost degenerate $3p_{3/2}$, $2f_{5/2}$ and 
$1i_{13/2}$ orbitals just below the Fermi level. A couple of MeV below
we find degenerate $1h_{9/2}$ and $2f_{7/2}$ orbitals. 
Those levels evolve with deformation and at $Q_{20}$ around
6.6 b a gap in the SPE spectrum signaling a region of low level density 
appears both in the proton and neutron spectra that is responsible for 
the prolate minimum observed in the axially symmetric potential energy 
curve (the minimum becomes a saddle point when the triaxial degree of 
freedom is considered). In the oblate side, at $Q_{20}$ = -6.6 b 
the neutron's Fermi level approaches another gap that is responsible
for the oblate minimum observed in the axially symmetric PEC. As discussed
below both minima are in fact saddle points as long as the $\gamma$ degree
of freedom is considered. In the case of $^{184}$W, the SPE spectrum
for neutrons show a gap near the Fermi level for $Q_{20}$ = 8 b. This fact
together with the proton's gap also observed in that region of $Q_{20}$
favors the development of the prolate minimum observed. In the oblate
side, both the neutron's and proton's SPE show no gap around the Fermi
level in the relevant range of deformation justifying the lack of such
a minimum. For the nucleus $^{196}$W we observe how the neutron's SPE
spectrum shows a gap around the Fermi level for oblate deformations
with $Q_{20}$ in the range between -1 b and -10 b that is responsible for
the oblate minimum observed in this case. Therefore the prolate-oblate
transition seen at N=116 is a consequence of the two gaps in the neutron's
SPE, one in the prolate and the other in the oblate side as the Fermi 
level crosses them. On the other hand, the proton's SPE spectrum seems
to favor the appearance of coexisting oblate and
prolate configurations as Z increases that are the precursors of the
triaxial instability observed in that case. 

We can also look at the onset of deformation in this region by using the
ideas developed by Federman and Pittel (FP) \cite{Federman.77} in trying to
unify the description of deformation both for light nuclei and heavy ones.
A recent study using the same ideas has been performed in Ref. \cite{Fossion.08}.
in the rare earth region. The argument of Ref. \cite{Federman.77} is that 
deformation is driven by the $T=0$ neutron-proton interaction and this is 
particularly intense between spin orbit partners. Next in the range of 
relevance of the n-p interaction strength we find interactions between 
orbitals with the same radial quantum number (FP's argument is written 
in the language of spherical shell model orbitals) and large orbital 
angular momenta differing by one unit (i.e., $n_p=n_n$ and $l_p=l_n\pm 1$). 
By looking at the SPE plots in Fig \ref{190Wspe} we find the relevant
role of the $1h_{11/2}$ orbital for protons which is very close to
the Fermi level for all the nuclei considered in the region. According to
FP's argument this orbital could interact with its neutron spin orbit
partner, namely the $1h_{9/2}$ orbital but this one is well below
the Fermi level and can be considered as inert. Near the neutron's Fermi
level we have a $1i_{13/2}$, $2f_{5/2}$ and $3p{3/2}$. Obviously, it is
the first one that fulfills the above criteria of $n_p=n_n$ and 
$l_p=l_n\pm 1 $ and therefore is the strongly interacting one with the 
$1h_{9/2}$ orbital. For values of N around 110 the $1i_{13/2}$ is in
the middle of the Fermi level favoring the observed prolate deformation
with well established and deep prolate wells. As N increases the $1i_{13/2}$
gets more and more occupied and at some point it ceases to play a role 
that is transferred to the $2f_{5/2}$ and $3p{3/2}$ orbitals. Among them,
only the $2f_{5/2}$ can interact with the $2d{3/2}$ of protons but as
the $l$ values are low we do not expect a strong interaction. This explains
why as N increases the depth of the deformation wells decreases favoring
triaxial deformations.

\begin{figure}
\includegraphics[width=0.95\textwidth,angle=0]{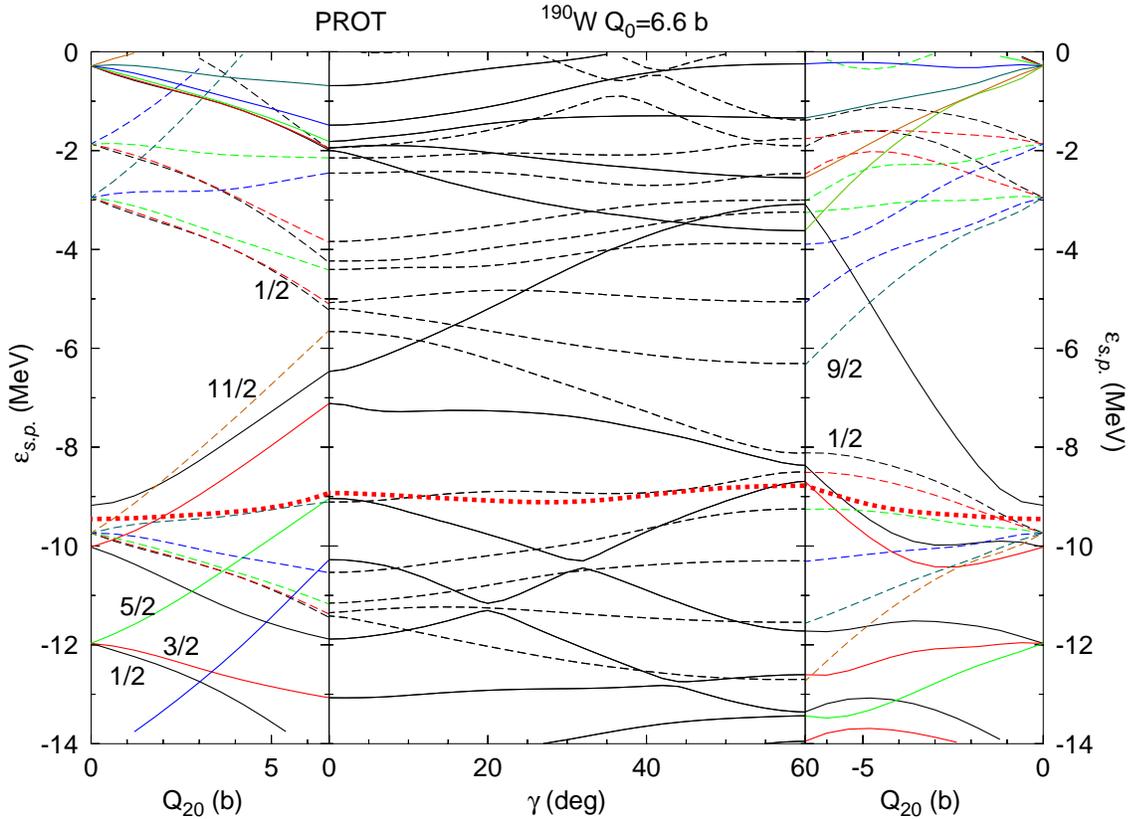}
\caption{(Color online) In this combined plot, the proton SPE for the nucleus $^ {190}$W are plotted.
In the left panel the axially symmetric SPE are plotted as a function 
of $Q_{20}$ from $Q_{20}=0$ up to $Q_{20}=6.6$b. In the
middle panel the triaxial SPE are plotted as a function of 
the $\gamma$ deformation parameter and for $Q_0=6.6$b (the position of the ground state
minimum). Finally, in the right-most panel, the axially 
symmetric SPE are plotted as a function of $Q_{20}$ from 
$Q_{20}=-6.6$b up to $Q_{20}=0$b. In the three cases the Fermi level 
is depicted as a thick dashed line. The results have
been obtained with the Gogny D1S force. Some $K$ values are 
given in the plot.}
\label{190WspeProttriax}
\end{figure}

\begin{figure}
\includegraphics[width=0.95\textwidth,angle=0]{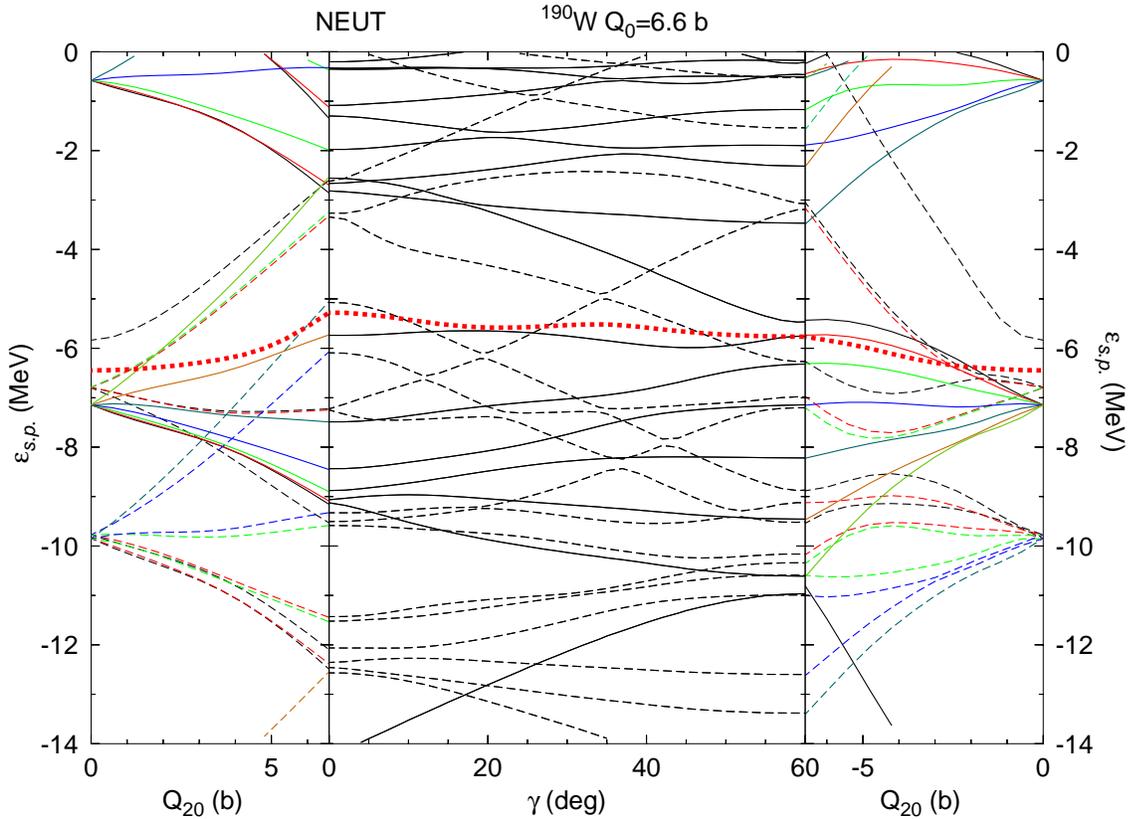}
\caption{(Color online) Same as in Fig. \ref{190WspeProttriax} but for neutrons.}
\label{190WspeNeuttriax}
\end{figure}

To further investigate the origin of triaxiality we have considered,
in addition to the axial SPE plots, also the single particle energies 
depicted as a function of the $\gamma$ degree of freedom and at 
a $Q_0 \equiv Q_{20}$ value of 6.6 b (that corresponds to the triaxial 
minimum)  for the $^ {190}$W nucleus. The triaxial SPE for protons
are depicted  in Fig. \ref{190WspeProttriax} whereas 
Fig. \ref{190WspeNeuttriax} is for neutrons. In those plots we have 
sticked together the SPE plots along the axially
symmetric $Q_{20}$ degree of freedom (leftmost panel for the prolate side, 
rightmost panel for the oblate side) with the SPE plots along the triaxial
degree of freedom $\gamma$ (middle panel). The main reason for this 
representation is to identify the $K$ values of the triaxial single particle 
levels at the axial limits corresponding to $\gamma=0 ^\circ$ and $\gamma=60 ^\circ$. 
The first fact worth mentioning is that the $K$ contents of most of the 
levels change as $\gamma$ evolves, in such a way that in most cases 
the $K$ value at $\gamma=0 ^\circ$ is
different from the $K$ value at $\gamma=60^\circ$. A typical example, in
the proton spectrum is the negative parity level with $K=1/2$ and located 
at $\sim -5.2$ MeV at $Q_0=6.6$ b and $\gamma=0$ that becomes at $\gamma=60 ^\circ$ 
degrees a $K=9/2$ orbital (originating from the same spherical subshell).

This is a direct consequence of $K$ mixing associated to the triaxial degree of freedom.
We also observe, both in the proton and neutron spectra several avoided level
crossings taking mainly place between $\gamma=15 ^\circ $ and $\gamma=45^\circ$.
Concerning the level density around the Fermi level we observe that the 
level density of protons is rather low around $\gamma=30 ^\circ$ and 
this fact is driving the system towards 
the observed triaxial minimum in $^{190}$W at this $\gamma$ value. 
On the other hand, the level density of neutrons remains rather 
high around the Fermi surface for the whole 
range of $\gamma$ values not favoring the development of a triaxial
minimum and indicating a more passive role of neutrons in 
the generation of triaxiality. We also notice that
the addition of extra protons (to have Os and Pt) will locate the Fermi 
level of protons in the middle of the observed gap (at this $Q_0$ value) 
driving the corresponding system (Os and Pt) to triaxiality as it is
observed as a general rule in the systematics of the $Q_0-\gamma$ planes
discussed previously. Also the less active role of neutrons in the 
development of triaxiality is consistent with the systematics of the 
$Q_0-\gamma$ planes as triaxiality seems to depend rather weakly on 
neutron number.

\subsection{Moments of inertia}

The moments of inertia of the first $2^+$ states have been computed for all the nuclei 
considered. The quantity computed is the Thouless-Valatin or first moment of inertia 
obtained by using the formula ${\cal J}^{(1)}=3/E_\gamma$, where $E_\gamma=E_{2^+}-E_{0^+}$ 
is the $\gamma$ ray energy for the $2^+\rightarrow 0^+$ decay. The theoretical 
energies involved in the previous definitions have been obtained using the 
selfconsistent cranking method (i.e., using in the HFB equations a time reversal breaking
constraint on the $x$ component of the angular momentum operator, 
$\langle \hat{J}_x \rangle =\sqrt{I(I+1)}$, which involves a Lagrange term 
of the form $-\omega \hat{J}_x$, see Ref. \cite{Egido.93} for an application with the
Gogny force). For the calculations we have used the
Gogny D1S and D1N forces. The reason 
for this choice is the success of the D1S parametrization in the description 
of many high spin-properties over the whole nuclide chart as well as
the scarce number of results available for SLy4.  The other parametrization
has been chosen because its pairing properties are slightly different 
from the D1S ones and therefore a comparison of the D1S and D1N moments
of inertia, which strongly depend upon pairing, can give a hint on the range of values where one can expect
a reasonable prediction. The results obtained are presented 
in Table \ref{TableMOMI} along with some experimental numbers extracted from the 
$E_{2^+}$ experimental energies. 
First of all, the D1S and D1N results are quite similar, showing a tendency
of bigger values for D1S as a consequence of its slightly reduced pairing
correlations as compared to those of D1N. The more pronounced differences
are due to slightly different values of the $\gamma$ deformation parameter
for the $J=0$ ground state. The coincidence of the results give us
confidence on the robustness of our theoretical predictions with respect to
a change in the interaction. Turning now to the results, they indicate 
an increase of the moment of inertia in going from neutron number N=110 to N=112 in 
the lighter isotopes Yb, Hf, and W as a consequence of the quenching of neutron pairing
correlations. This effect is not observed in the experimental data. From N=112  
and up to the maximum neutron number 
considered, the moments of inertia decrease as corresponds for a decreasing 
deformation parameter $\beta$ (exceptions are the N=116 Hf and W isotopes;
they correspond to the Hf and W isotopes where the onset of triaxial 
deformation takes place, see Table \ref{TableBetaG}). 
Regarding the comparison with the experiment we observe that the 
selfconsistent cranking results tend to overestimate the experimental 
values. This is a well known effect, consequence of too low pairing 
correlations at the mean field level. The cure to this deficiency 
implies the use of beyond mean field techniques in the treatment of 
pairing correlations (mainly by  restoring the number of particles 
using projection techniques) which is out of the scope 
of this paper. Finally, let us conclude this section with the following
remark: the values of the moments of inertia obtained do not show any
significant and systematic differences when the ground state of the corresponding nuclei
are axially symmetric or triaxial. We conclude that the moment of inertia
is not a good quantity to disentangle the character of the ground state 
deformation of the nuclei in this region.

\begin{table}
\begin{tabular}{|c|c|c|c|c|c|}\hline
N   &   Yb   &    Hf          &       W        &       Os        &    Pt   \\ \hline\hline
110 & 35.97  & 32.50 (30.7)   &  29.24 (27.0)  &   30.86 (21.9)  &   25.96 \\ 
    & 36.09  & 31.76          &  29.17         &   30.17         &   25.17 \\ \hline
112 & 43.83  & 35.60 (27.9)   &  30.71 (24.5)  &   30.67 (19.4)  &   25.08 \\ 
    & 39.41  & 35.56          &  31.81         &   29.92         &   23.81 \\ \hline
114 & 34.83  & 28.16          &  25.38 (21.0)  &   30.29         &   24.72 \\ 
    & 30.19  & 27.51          &  25.34         &   29.38         &   23.00 \\ \hline
116 & 28.92  & 29.46          &  28.87         &   27.49         &   23.01 \\ 
    & 28.72  & 28.35          &  26.85         &   25.80         &   20.36 \\ \hline
118 & 23.28  & 24.61          &  25.73         &   25.17         &   19.33 \\ 
    & 23.09  & 24.32          &  23.70         &   23.04         &   16.45 \\ \hline
120 & 20.55  & 17.33          &  22.05         &   20.68         &   13.19 \\ 
    & 15.43  & 21.48          &  21.57         &   19.70         &   12.23 \\ \hline
122 & 10.15  & 10.44          &  10.52         &   10.31         &    9.56 \\ 
    & 10.05  & 10.17          &  10.05         &    9.69         &    9.08 \\ \hline
\end{tabular}
\caption{Static moments of inertia ${\cal J}^{(1)}$ (in MeV$^{-1}$) for the first $2^+$ rotational
states obtained with the Gogny D1S force (upper rows) and D1N (lower rows) 
and the selfconsistent cranking method. In parenthesis, in the upper rows
the experimental results for those nuclei with a ratio $E_{4^+}/E_{2^+}>3$ as to make sure
that they are reasonable rotors.}\label{TableMOMI}
\end{table}

\section{Conclusions}

We have presented the results of triaxial mean field calculations for several
isotopes of the Yb, Hf, W, Os and Pt nuclear species with neutron numbers
ranging from N=110 up to N=122. The aim is to explore how the ground state
deformation evolves in these nuclei. In order to establish in firm 
grounds the validity of our findings we have performed the calculations
with two different parametrizations of the Gogny force, D1S and D1N, and
with the SLy4 parametrization of the Skyrme energy density functional. Those
forces/functionals differ in the range of their central parts as well as
in their pairing properties and therefore it is to be expected that 
nuclear deformation characteristics depending upon tiny details of the
force/functional will differ in the various calculations. On the other hand,
common characteristics present in the two types of calculations can be considered
as force/functional independent and therefore as more robust predictions.

We have shown that increasing the proton number in this mass region
leads the nuclei to triaxiality. On the other hand, increasing the neutron
number, the ground state shapes in the isotopes studied evolve from
axially deformed prolate shapes to axially deformed oblate shapes. The
transitional nuclei (N$\approx$116) exhibit a $\gamma$ soft behavior
with very shallow triaxial minima. The transition occurs with different
degrees of stiffness depending on the isotope.
The transition is rather sharp for the low Z isotopes Yb and Hf but 
is much broader for W, Os and Pt where a region of triaxial ground states
develops in between the region of prolate and oblate minima. Several
isotopes of W, Os and Pt develop triaxial minima but their depths, which
are rather low in general, depend strongly  on the interaction/functional
considered. For this reason, we can only conclude that triaxial effects
will surely play a role in the above mentioned cases but the extent to
which they influence the nuclear spectrum is still uncertain and calculations
considering fluctuations on the deformation parameters (Bohr hamiltonian-like)
are needed.

The analysis of the single particle energies both for axially symmetric
and triaxial configurations demonstrates the role of different gaps showing up
in the SPE of both protons and neutrons as well as the role played by
the $T=0$ proton-neutron interaction. Concerning the driving force towards
triaxiality, we can conclude that in this region protons play a more relevant
role than neutrons.

Finally, the comparison of the selfconsistent moments of inertia shows that this
is not the right quantity to look at in order to disentangle the characteristics
of the ground state deformation of the nuclei in this region.

\section*{Acknowledgments}
Work supported in part by MEC, Spain (FPA2007-66069) and MICINN, Spain (FIS2008-01301) 
and by the Consolider-Ingenio 2010 program CPAN (CSD2007-00042)

\end{document}